# Considering Durations and Replays to Improve Music Recommender Systems

Pierre HANNA

Computer Science Laboratory (LaBRI), University of Bordeaux, Cours de la Libération, 33405 TALENCE, FRANCE
Simbals SAS, 27 Allée des Petits Rois, 33400 TALENCE, FRANCE

pierre.hanna@labri.fr

## ABSTRACT

The consumption of music has its specificities in comparison with other media (movies, books), especially in relation to listening durations and replays. Music recommendation can take these properties into account in order to predict the behaviours of the users. Their impact is investigated in this paper. A large database was thus created using logs collected on a streaming platform, notably collecting the listening times. The proposed study shows that a high proportion of the listening events implies a *skip* action, which may indicate that the user did not appreciate the track listened. Implicit *like* and *dislike* can be deduced from this information of durations and replays and can be taken into account for music recommendation and for the evaluation of music recommendation engines. A quantitative study as usually found in the literature confirms that neighborhood-based systems considering binary data give the best results in terms of MAP@k. However, a more qualitative evaluation of the recommended tracks shows that many tracks recommended, usually evaluated in a positive way, lead to *skips* or thus are actually not appreciated. We propose the consideration of implicit *like/dislike* as recommendation engine inputs. Evaluations show that neighbourhood-based engines remain the most precise, but filtering inputs according to durations and/or replays have a significant positive impact on the objective of the recommendation engine. The recommendation process can thus be improved by taking account of listening durations and replays. We also study the possibility of post-filtering a list of recommended tracks so as to limit the number of tracks that will be unpleasantly listened (*skip* and implicit *dislike*) and to increase the proportion of tracks appreciated (implicit *like*). Several simple algorithms show that this post-filtering operation leads to an improvement of the quality of the music recommendations.

## 1 Introduction

Facing the mass of digital data available today, individuals do not have sufficient personal experience or competences to be able to make their own choices. Recommender systems[1] are taking an increasingly important part in the digital applications currently being developed and proposed to the public: e-commerce, news articles, music, video, book, ... They help users make their decisions by providing lists of items that should be most likely of interest to them. The design of recommendation engines has thus been the subject of numerous scientific researches, compiled in several books[2,3].

Different types of recommender systems have been developed[4]. systems based on collaborative filtering are historically the first recommender systems[5]. They take into account the interactions between users and items, in order to highlight associations between users and particular items that these users have not yet consumed. Thus, a recommended item is an item liked by a user with similar tastes. In a different way, content-based systems[6] consider both descriptors of the content of items and positive or negative ratings made by users for these items. By considering similarity based on the content, new items similar to those previously appreciated can be provided to each user. Finally, hybrid recommender engines[7] combine different types of engines to balance the advantages and drawbacks of each technique.

Collaborative filtering approaches can be grouped in the two general classes of neighbourhood and model-based methods. Neighbourhood-based (or memory-based) collaborative filtering methods consider user ratings stored in memory to predict ratings for new users or new items[8]. Two approaches can be applied, taking into account similarities between users (user-based recommendation[9]) or similarities between items (item-based recommendation[10,11]). These approaches therefore process ratings to compute predictions, while model-based approaches use these ratings to learn a predictive model[12,13]. This model captures the important elements of relationships between users and items, learning the parameters of the models from an existing dataset.

Whatever the approaches considered, they are therefore based essentially on the ratings of items provided by users. These ratings can be explicit, i.e. directly made by users, or implicit[14], i.e. deduced from the actions of users on items or on the browsing for items . Examples of explicit ratings include the Netflix Video Streaming Platform, which asks for its users to rate the movie they had just watched by providing 1 to 5 stars. This explicit action is binding for some users. For example, Netflix has just suspended its explicit rating method used for several years to propose a less restrictive rating process [1]. The Deezer or Spotify music streaming platforms also allow their users to indicate their satisfaction by clicking on a *thumb-up* or a *thumb-down* on each track listened or to select a track for their *loved track* list.

Many existing recommender systems deal with explicit ratings (e.g. Netflix). However, this explicit data requires additional user commitment which may appear to be binding. As a result, the number of explicit ratings is often limited. Moreover, explicit ratings can sometimes be misleading because they are often accessible to other users: a user who explicitly expresses *I love this track* shows the community his or her tastes, whereas he or she does not always want to share them.

For each user, numerous implicit usage parameters, such as clicking, listening or browsing histories, are considered in order to precisely describe its usage and may thus be taken into account by the recommender systems in order to estimate its preferences. For example, a user who listens to several tracks from the same performer probably likes this performer and may appreciate other tracks from that same performer.

Therefore, other existing recommender systems deal with implicit data, such as consumption history or browsing history[15]. It is then necessary to determine the most pertinent descriptors of implicit ratings. Implicit positive ratings are often rightly considered since the choice of an object by a user is a strong implicit rating. But the possibilities to induce negative ratings from user actions are still under-exploited, such as an interrupted film or an interrupted book, comments with negative words, etc. Indeed, these implicit ratings are inevitably noisy because they are deduced indirectly from user actions. Another difficulty inherent in recommender systems based on implicit rating is the evaluation of these systems. The estimation of explicit ratings is a clear goal and is discussed in the literature[3]. It can be evaluated directly. The evaluation of a system of estimation of implicit ratings is more debatable.

Among the applications of recommender systems, music recommendation is very important in the context of digital music streaming: Spotify, Deezer, Google, Apple, Amazon, Rhapsody, ... Many systems of music recommendation have been listed in surveys[16]. The consumption of music exhibits specificities that must be taken into account by the recommendation engines. For example, facing the large amount of tracks available (several tens of millions on some platforms), recommender systems can consider music at the artist level[17], at the album level or at the track level.

The study presented in this article is devoted to music recommendation, choosing to place itself at the musical track level. The item considered here is therefore a musical track. No descriptors of music tracks are taken into account : no content-based approach is applied. Consumption data considered concerns a short period of time in order to reduce effects of temporal dynamics on user tastes[18].

Generally, existing music recommendation engines take as input the data corresponding to the collection of the clicks of users on tracks. Sometimes, these clicks are filtered in order to take into account only track streams that last at least 30 seconds. However, a click or a 30-second play does not necessarily imply a complete listening of the track. It also does not systematically mean that the user likes the track. Indeed, a listening can be passive, such as listening to a radio or a playlist. Otherwise, it can be active, with the explicit choice of a track by the user. Thus, when listening is passive, listening does not even correspond to a choice of the user. Moreover, if the user launches a playlist or a radio, he can passively listen to a track or a short part of a track which he does not like and which he would not have liked to be suggested.

These peculiarities of the music consumption lead to consider other meaningful elements among the usage data to supplement the explicit data and to deduce from them implicit ratings. Early studies indicate that implicit ratings may be more accurate in the area of the music recommendation[19]. Browsing and listening histories seem particularly relevant to describe the rating of a track by users. Indeed, a user who particularly enjoys a track will have a strong tendency to listen to it again. On the other hand, a track that has not be appreciated will probably not be listened again. Similarly, a short time listening tends to indicate that the user has not wanted to listen to the track until the end, what can be translated by the fact that he does not appreciate it too much. Thus, taking into account these implicit rating data may improve the quality of the recommendations. Early work on listening for the recommendation based on collaborative filtering have led to promising results[20].

However, more recent results, such as the Million Song Dataset[21] (MSD) challenge, have shown that the most accurate systems have been obtained by neighbourhood-based recommender systems with binary data as input[22]. Taking into account the play counts did not have a positive impact on the accuracy of the systems evaluated.

The study proposed in this paper deals with two issues raised by the various observations relating to music recommendation:

1. Is it possible to improve the relevance of the recommended tracks by taking account of durations and replays?

2. From a list of music tracks, is it possible to obtain a precise estimation of the ratings of these tracks by a specific user? Is it possible to filter out the tracks that this user would probably not appreciate?

To address these two problems, the data collected for the proposed experiments are presented in Section 2.1. An analysis of the users' behaviour with respect to the durations and replays is presented in Section 2.2. Then, the experimental protocols and the results of evaluations of difference recommendation algorithms are presented for different recommendation algorithms in Section 3. In particular, different input data are compared in Section 3.3 and several filtering algorithms are proposed and tested in Section 3.4.

---

[1]https://media.netflix.com/en/company-blog/goodbye-stars-hello-thumb

## 2 Analysis of the dataset

In this section, the dataset used for the experiments of this article is described. Durations and repetitions of music plays are studied in order to deduce criteria of implicit *like/dislike*.

In the following, the set denoted by $U$ represents the set of users. The set denoted by $I$ represents the set of tracks. The set denoted by $E$ represents the set of listening events (assumed as a user has listened to a track). Each event is related to a user, a song and a time-stamp $t$. They are therefore denoted by $E(u,i,t)$ with $u \in U$ and $i \in I$.

### 2.1 Dataset

One of the references concerning database for recommender system experiments is the dataset proposed by Netflix[23]. This dataset consists of approximately 1 million explicit ratings (1 to 5 stars) of over 17,000 movies by roughly 500,000 users. The algorithm competition proposed with this dataset consisted of comparing prediction algorithms. In the area of movie recommendation, other very high size datasets, such as MovieLens[24], also allow comparisons of recommender systems.

In the field of music recommendation, several evaluation datasets have been proposed. Based on the Netflix challenge, the KDD cup[25] was organized in 2011 and is based on data from the Yahoo Music service. The challenge was also to estimate user preferences from explicit ratings. The amount of data is very huge, with over 250 million ratings available. Another dataset, named *Million Song Dataset*, is one of the most widely used in the field of musical information retrieval. The tags and audio features of a million songs of over 40,000 artists are made available. The challenge *MSD*[21], organized in 2012 and mostly based on this dataset, consisted of comparing the prediction algorithms of the listening events. More recently, the dataset *LFM-1b*[26] delivers a collection of one billion songs by 120,000 users. The peculiarity of this dataset comes from the availability of the timestamps of each listening event. The dataset *30Music*[27] makes available roughly 31 million listening events by over 45,000 users, with the particularity of grouping listening sessions.

In this paper, a new dataset specific to music is considered. The originality concerns the availability of duration of the listening events, mostly not proposed by other datasets. Thus, the duration of the listening events enables to know whether the tracks have been listened to in their entirety or not. By way of comparison, no *skip* events, e.g. very short listening events, are available in the *30Music*[27]. For the study proposed, the data was collected on an international music streaming platform during the month of September 2016. This limited time over a month is deliberately short to avoid the effects of dynamic on the evolution of musical tastes [28]. The anonymised data is composed of over 180 million listening events of over 2,850,000 tracks by roughly 450,000 users. Table 1 shows more details about the dataset. The dataset has not been filtered, which implies that some listening events result from passive sessions such as radios or playlists, while other events result from an explicit choice of users. Minimal activity was required for user selection, minimizing the possibility that these users would use other music sources[29].

| Data | Tracks $|I|$ | Users $|U|$ | Events $|E|$ | unique pairs $(u,i)$ |
|---|---|---|---|---|
| Data $A$ | 2,442,392 | 383,678 | 100,000,000 | 54,390,720 |
| Data $B$ | 2,004,193 | 87,785 | 80,255,408 | 30,776,280 |
| Total | 2,850,118 | 471,463 | 180,255,408 | 85,167,000 |

**Table 1.** Details about the two parts of the dataset used for recommender system experiments .

Following the model of the *MSD* challenge, the collected data are separated into two groups. The group $A$ is limited to 100 million listening events. The group $B$ is composed of over 80 million listening events by 87,785 users. In this group $B$, for each user, the listening events are separated into two subgroups: *visible* events (subgroup denoted by $B_V$) and *hidden* events (subgroup denoted by $B_H$). Concerning the study presented in this paper, data consists of to quadruplets <user, track, duration, time>. The set of these quadruplets is denoted by $E$. The data of the groups $A$, $B$, $B_H$ and $B_V$ are denoted respectively by $E_A$, $E_B$, $E_{B_V}$ and $E_{B_H}$. $E(u,i,t)$ represents the quadruplet of $E$ whose user is $u$, whose track is $i$, and whose timestamp is $t$:

$$E(u,i,t) = (u,i,d,t) \in E, \forall d \qquad (1)$$

Data are anonymised, so tracks and users are represented by integer identifiers. The duration time is expressed in seconds and is denoted by $d(u,i,t)$ for the duration of the listening event $E(u,i,t)$ of the track $i \in I$ by the user $u \in U$, at the time $t$.

## 2.2 Quality of listening events

The dataset used for this study allows to attribute to each interaction between a track and a user the duration of listening. On the other hand, it is also possible to deduce directly from these data the repetitions of listening, e.g. the replays, that is to say the number of times that a user listens to the same track. Before considering these two properties as input to the recommender systems, it is necessary to evaluate their distribution.

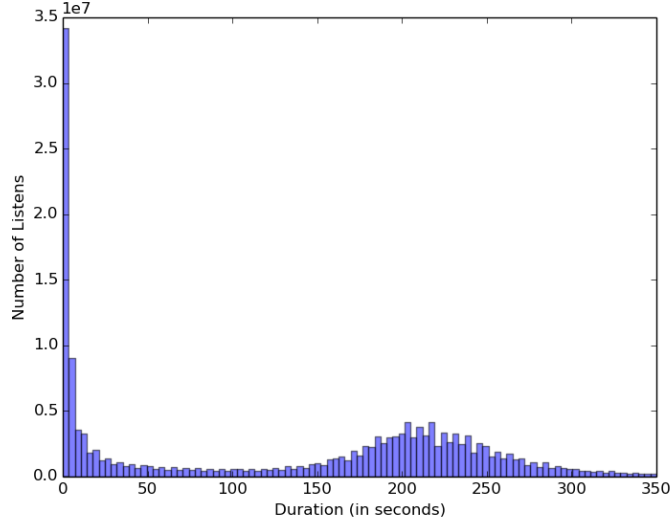

**Figure 1.** Distribution of the listening times (in seconds) in the dataset.

| Data | Events $< 5s$ | Events $< 30s$ | Events $< 60s$ | Events $< 120s$ |
|---|---|---|---|---|
| Data A | 24,424,226 (24.4%) | 34,103,711 (34.1%) | 37,842,100 (37.8%) | 42,934,539 (42.9%) |
| Data B | 19,461,415 (24.2%) | 27,248,930 (34.0%) | 30,305,101 (37.8%) | 34,519,326 (43.0%) |

**Table 2.** Distribution of the durations of listening events.

The listening durations of the tracks vary widely, ranging from one second to the full duration of the track. Figure 1 presents the distribution of these listening durations. Table 2 gives some quantitative data, e.g. values of $|\{d(u,i,t) \setminus d(u,i,t) < d_0\}|$ pour $d_0 \in \{5, 30, 60, 120\}$ seconds.

The first observation is the high proportion of very low durations: almost a quarter of the listening events lasts for less than 5 seconds, nearly two thirds of the listening events last for more than 30 seconds. These short durations can be explained by several behaviours. For example, some users listen lists of tracks (for example, playlists) by clicking and listening only a few seconds each a song in order to find a particular track. Other users listen to songs passively from a themed radio, but perform regular skips on tracks they do not like. The distribution of the durations beyond the 30 first seconds indicates that after a minimum duration of listening, users reach more and more the end of the track listened to. Since the duration of a track is not always the same, a uniform distribution of the duration of high values can be expected. Yet it appears on Figure 1 a Gaussian type distribution with a greater proportion of listening durations around 210 seconds. This is justified by the greater proportion of popular existing tracks whose duration is around 3 minutes and 30 seconds, the favourite format for singles and radio broadcasts.

Replays of the track $i$ by the user $u$ are denoted by $p(u,i)$, for $(u,i) \in U \times I$, and defined according to the following equation 2 :

$$p(u,i) = |\{E(u,i,t), \forall t\}| \qquad (2)$$

Analysis of the distribution of these replays $p$ also reveals very different behaviours. Figure 2 shows the distribution of these replays. Table 3 gives some quantitative data by presenting the number of replays greater than 2, 5 and 10, that is to say $p_{p_0} = |\{p(u,i) \setminus p(u,i) \geq p_0\}|$ for $p_0 \in \{2, 5, 10\}$.

Finally, analysis of the listening durations reveal a high proportion of durations with low values. However, either these short-length plays are not taken into account in the existing studies on the music recommendation, or the listening events are taken into account in the same way, whatever their durations. Similarly, replays represent a significant proportion of listening events. In order to consider these usage properties in a recommendation engine, it is necessary to translate them into implicit ratings.

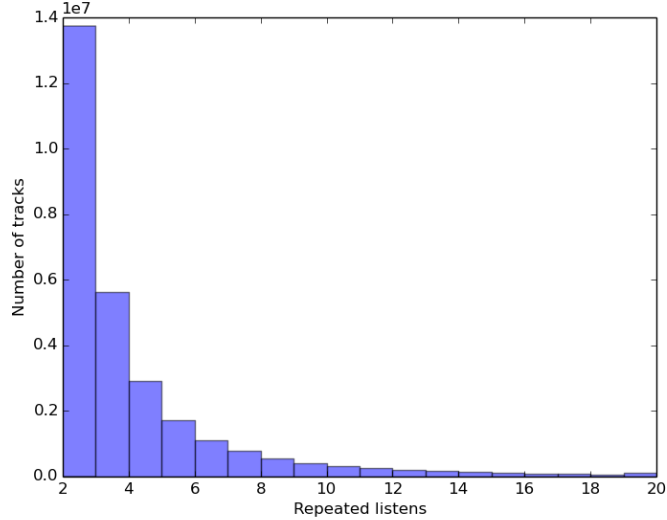

**Figure 2.** Distribution of replays in the dataset.

| Data | $p_2$ | $p_5$ | $p_{10}$ |
|---|---|---|---|
| Data $A$ | 16,757,644 | 3,156,017 | 817,203 |
| Data $B$ | 12,196,295 | 3,535,469 | 1,282,051 |

**Table 3.** Distribution of replays $p_{p_0}$ for $p_0 \in \{2, 5, 10\}$.

### 2.3 Towards implicit *like/dislike*

For applications related to recommendation, explicit ratings are often scores, such as in the Netlfix system, or sometimes just binary ratings to indicate whether the product is loved by the user (for example a heart or a thumb-up) or, on the contrary, is not loved by the user (eg a thumb-down or a crossed heart). Obviously, this way of indicating a rating implies that all the items are not rated, since the user only submit a rating for the tracks for which his opinion is definitive. In the following, the terms of *like* (the user likes the track) and *dislike* (the user does not like the track) correspond to this binary rating.

However, these explicit *like/dislike* ratings are not always available in applications. Moreover, even when this is the case, they are not always provided by the users because they are often perceived as binding. They can also be distorted by the fact that they are sometimes shared with other users. This is why it is experimented here to deduce from the interactions of the user implicit ratings under the *like/dislike* form. Several definitions are proposed:

- Listening event is assumed as short, e.g. related to a *skip* action, when a track has been listened to, but the listen duration has been less than 30 seconds. This shortened listening of a track $i$ by user $u$ at time $t$ is denoted by $K(u,i,t)$ and is defined by:
$$K(u,i,t) = \{E(u,i,t) \setminus d(u,i,t) < 30\} \tag{3}$$
$K(u,i)$ is the number of shortened plays : $K(u,i) = |\{K(u,i,t), \forall t\}|$.

- A listening event is named as a *stream* if it lasts at least 30 seconds, which usually corresponds to the market standard. This listening event, whose duration is over 30 seconds of a track $i$ by user $u$ at time $t$ is denoted by $S(u,i,t)$ and is defined by:
$$S(u,i,t) = \{E(u,i,t) \setminus d(u,i,t) \geq 30\} \tag{4}$$

- Replay is assumed when a track is listened for more than 30 seconds by the same user. The number of listening events for track $i$ by user $u$ whose duration is greater than 30 seconds is denoted by $P(u,i)$ and is defined by:
$$P(u,i) = |\{S(u,i,t), \forall t\}| \tag{5}$$

On the basis of these definitions, as well as the previous observations of the duration and replay distributions, a definition for *like* and *dislike* implicit ratings is proposed for the following:

- **Implicit *like*** : a user $u$ has listened at least twice to the same track $i$ more than 30 seconds, e.g. if $P(u,i)$ $geq 2$, without any skip event, e.g. $K(u,i) = 0$. This choice is justified by the intuition that replaying a track over a long time suggests that the user is satisfied with this track. In this case, implicit *like* rated by the user $u$ for the track $i$, denoted $L(u,i)$, is 1. Otherwise, it is 0.

$$L(u,i) \longrightarrow \begin{cases} 1 & \text{if } P(u,i) \geq 2 \text{ and } K(u,i) = 0, \\ 0 & \text{else.} \end{cases} \tag{6}$$

- **Implicit *dislike*** : a user $u$ has skipped at least one time a track $i$, e.g. $K(u,i) > 0$, without having ever listened to it another time, e.g. $P(u,i) = 0$. This choice is justified by the intuition that shortening a play indicates that the related track is not appreciated. In this case, implicit *dislike* rated by the user $u$ for the track $i$, denoted $D(u,i)$, is 1. Otherwise, it is 0.

$$D(u,i) \longrightarrow \begin{cases} 1 & \text{if } K(u,i) > 0 \text{ and } P(u,i) = 0, \\ 0 & \text{else.} \end{cases} \tag{7}$$

It is important to note that a track $i$ on which a user $u$ interacts does not always imply an implicit *like* or *dislike*. In the definitions proposed, a decision is not always taken. Thus it is possible for a couple $(u,i)$ that $L(u,i)$ and $D(u,i)$ are both 0. On the other hand, for a couple $(u,i)$, it is impossible for $L(u,i)$ and $D(u,i)$ to be both 1:

$$\begin{aligned} L(u,i) = 1 &\Rightarrow D(u,i) = 0 \\ D(u,i) = 1 &\Rightarrow L(u,i) = 0 \end{aligned} \tag{8}$$

These choices for implicit *like* and *dislike* definitions are debatable but appear justifiable by observations of average usage. Obviously, each user behaves differently, and the ideal would be to automatically adapt these definitions to each user. The experiments described below are not intended to validate these choices, but to study the impact of these very simple criteria on the quality of the recommendations. Table 4 gives the number of tracks concerned by streams, skips, implicit *like* and *dislike*. It is important to note that their number and ratio are significant, notably the quantity of tracks with implicit *dislike* (around 30%). In the following, the objective is to try to improve the quality of the music recommendations by reducing the numbers of skipped tracks or implicit *dislike*, while maximizing the numbers of streams and implicit *like*.

|  | Data *A* | data *B* |
|---|---|---|
| Streams $S(u,i)$ | $65,896,289$ (65.9%) | $53,006,478$ (66%) |
| Implicit like $L(u,i) = 1$ | $7,217,823$ (13.3%) | $2,373,467$ (15.4%) |
| Skips $K(u,i)$ | $34,103,711$ (34.1%) | $27,248,930$ (34.0%) |
| Implicit dislike $D(u,i) = 1$ | $17,062,682$ (31.4%) | $4,584,936$ (29.7%) |

**Table 4.** Distribution of the different types of listening events in the experimental datasets.

## 3 Qualitative study of recommendations

From the dataset described above, a music recommender system can be developed and evaluated. Several types of models can be compared according to the usual criteria, such as the metrics used in the *MSD* Challenge[21], but also by taking into account the implicit rating criteria proposed.

### 3.1 Algorithms et implementations applied

Several types of recommendation engines are proposed in the literature[3]. However, content-based approaches are not possible for the available test database. Collaborative filtering approaches can be tested using dedicated implementations such as *Apache Spark*[30] or *Apache Mahout*[31].

*Apache Spark* is a library that contains an implementation of a model-based method that allows the modelling of the links between tracks and users by latent factors. These factors are computed by applying the algorithm *alternating least squares* (ALS)[32]. This implementation has the advantage of being able to consider as input explicit or implicit data. When the data are implicit, these are considered as confidence levels[15]. The comparison between these two explicit or implicit approaches is thus made possible. *Apache Mahout* also offers several implementations of different types of recommendation algorithms. The popular SVD (Singular Value Decomposition) algorithm allows to obtain latent factor modelling, thus estimating user ratings for tracks[33]. This approach has proven to be one of the most accurate within the Netflix prize[23].

Model-based approaches prove to be accurate for a goal of predictive estimation of ratings of tracks by users. However, for the purpose of finding a list of tracks that a user will appreciate to listen to, the results of the MSD challenge have shown the limitations of model-based approaches. *Apache Mahout* also provides implementations of neighbourhood-based (user-oriented or item-based) methods. Since the ratio between tracks and users in the dataset makes the application of the algorithms based on the neighbourhoods between items too time consuming, only the algorithms based on the neighbourhoods between users are experimented. However, tests of item-based algorithms that we carried out on limited datasets yielded results similar to those obtained by applying user-based algorithms.

Model or neighbourhood-based approaches take as inputs explicit or implicit ratings of a track by a user. These ratings are number values (integers in general). From the dataset proposed here, it is possible to consider the number of listening events $P(u,i)$ for each pair $(u,i) \in U \times I$, or only one binary value describing the fact whether the user has listened (possibly considering a minimum listening time) the track or not. Another approach tested here in Section 3.3 is to reduce the usage data describing the interaction of a user on a track to get only one rating score. This score is implicit and can then be input of algorithms.

### 3.2 Evaluation of recommendation engines

Evaluation of recommender systems is a difficult problem[34]. Here, the evaluation proposed is offline. It evaluates the ability of the recommender system to retrieve the tracks the user will listen to. A part of the tracks listened to by this user being known, the objective is to find the list of the other tracks that the user has actually listened to. The evaluation is therefore based on a hidden part of the dataset, following the process of the challenge MSD. To do this, two groups of users of the dataset were randomly selected. From the first group, the data *A* relating to these users are grouped together and are entirely available for learning models or neighbourhoods. From the second group, the data *B* are constituted. This data *B* is split into two parts: for each user, a random half of the listening events constitutes the data $B_V$, visible for learning, while the other hidden half constitutes the data $B_H$, remaining only available for evaluation. It was chosen to consider only one fold, since the time necessary to compute recommendations is very high and since the number of users in the dataset is very high. The set of users of the dataset *B* is denoted by $U_B$.

Many evaluations of recommender systems, such as the Netflix prize, rely on Root Mean Squared Error (RMSE) metrics, measuring the similarity between predicted and actual estimates given by users. Since the objective of the evaluation is to retrieve the tracks listened by each user, the evaluation metrics are those used in evaluations of information retrieval systems. Mean Average Precision at K (MAP@k) is measured following the evaluation computed during the MSD challenge[35]. The average precision is calculated at the rank *k*:

$$A\_P_k(u) = \frac{\sum_{m=1}^{k} \text{Prec}(m,u) \times \text{rel}(m,u)}{|E_{B_H}(u)|} \qquad (9)$$

where Prec$(m,u)$ represents precision at position *m* in the list of tracks recommended for user *u*, rel an indicator function equaling 1 if the track at rank *m* has been listened by user *u*, 0 otherwise. $E_{B_H}(u)$ is the set of *unique* tracks with which the user *u* has interacted, in the hidden data $B_H$ of the group *B*. From the average precision at rank *k*, the mean of the average precision, MAP@k, is calculated from:

$$\text{MAP@}k = \frac{\sum_{u \in U_B} A\_P_k(u)}{|U_B|} \qquad (10)$$

Table 5 presents the results obtained by running different algorithms. The results of an algorithm recommending the most popular tracks are proposed for comparison. Some algorithms take binary data as input, others listening event counts. The implementations named by *ALS explicit* and *ALS implicit* rely on the ALS algorithms implemented in Spark library, with explicit or implicit ratings respectively. Parameters were empirically chosen, under the constraints of the time computation: 50 latent factors, 20 iterations, 0.07 as regularization parameter. The implementation named by *Mahout User-Based Binary* rely on the function *GenericBooleanPrefUserBasedrecommender* from the Mahout library. It is based on the neighbourhoods between users (100 users), calculated on binary data with Tanimoto distance. The best results are obtained by the algorithm

*Mahout User-Based Binary.* The difference with recommendation by popularity is significant. The resulting values are similar to the results obtained during the challenge MSD[22]. Model-based methods, either SVD or ALS, give less accurate results, and even worse than recommendation based on popularity. It is important to note that results are less accurate with implicit data (the number of listening events here) than with binary data. Music recommender system based on binary rating is evaluated as more accurate comparing to systems based on explicit data. This result has already been observed[36].

| Algorithmes | MAP@10 | MAP@100 | MAP@500 |
|---|---|---|---|
| Popularity | 0.15307 | 0.05102 | 0.03891 |
| Mahout User-Based binary | **0.48216** | 0.21373 | 0.16305 |
| Mahout SVD | 0.01771 | 0.00371 | 0.00273 |
| Spark ALS explicit | 0. | 0. | 0. |
| Spark ALS implicit | 0.36724 | 0.16353 | 0.12310 |

**Table 5.** MAP@k (k=10,100,500) for different recommendation engines.

Finally, the results show that given the list of tracks listened a user, a recommender system based on binary data find more precisely the tracks that the user will then listen to than a system based on popularity. This evaluation criterion cannot be considered exhaustive since it concerns the tracks that the user will listen to, even without using a recommender system. The recommendation of these tracks is however of great importance. Indeed, proposing to a user the tracks that he would listen to reassures him and thus favours a better acceptance[37].

However, a more qualitative analysis of recommended tracks shows limitations to good quantitative results. Tracks recommended and listened by users are not always complete listening event: 39% of listening events concerning recommended tracks last less than 30 seconds and are nevertheless evaluated positively. Similarly, listening event does not always mean a positive appreciation of the track : 21% of listening events match the implicit *dislike* criteria, for only 22% for the implicit *like* criteria. These limitations express the need to evaluate the quality of the tracks recommended, in order to obtain recommendations that will favour the tracks which will be appreciated by the users (*like*) and avoid the undesired or unappreciated tracks (*skip*, *dislike*). This is the purpose of the suggestions and experiments presented in the next two sections.

### 3.3 Improvement by considering durations and replays

The input data of the recommender systems are made up of all the tracks listened to by users, whatever the duration of the listening event. In order to answer the first question raised in Sectionsec:introduction, it is proposed here to filter the input listening data according to durations and replays. Three inputs are compared: all listening events (i.e. no filtering), durations over 30 seconds (i.e. streams) and implicit *likes*.

The three induced systems are evaluated according to MAP at ranks 10, 100 and 500. They can be compared with the reference recommender system based on the popularity of the tracks, whose evaluation results are presented in table 5. As the objective is to evaluate also the quality of the recommended tracks, 5 evaluations of MAP on 5 different criteria are proposed : listening events (MAP$_E$@k), streams (MAP$_S$@k), *like*(MAP$_L$@k), skips (MAP$_K$@k) and *dislike* (MAP$_D$@k). In equation 9, the cardinal of the set $E_{B_H}(u)$ in the denominator and the rel function are adapted to only consider respectively listening events relating to the set $E$, streams relating to the set $S$, *likes* relative to the set $L$, skips relative to the set $K$ or *dislikes* relative to the set $D$. The definitions of MAP@k are thus different for these 5 evaluations. For example, for the calculation of MAP@k considering *like*, only the recommended tracks that gave rise to an implicit *like* are considered positive, e.g. for user $u \in U_B$, the tracks $i$ such that $L(u,i) = 1$.

The expected objectives are therefore different according to the criteria. For listening events, streams and *likes*, the highest MAP@k value is expected, as it would indicate that the tracks listened to, appreciated or liked, are highly ranked in the list of the recommended tracks. On the contrary, for skips and *dislikes*, the lowest value of MAP@k is expected because it would indicate that short-time listened or disliked tracks are not highly ranked, or even absent, in the list of the recommended tracks. Table 6 shows the different MAP@k values obtained, using the recommendation engine based on the neighbourhoods between users, denoted UB (User-Based), from the binary listening data.

In general, MAP@k values are relatively low. But it is important to remember that no filtering (for example considering popularity) of tracks is processed during our experiments. Nevertheless, MAP@k values remain much higher than those obtained with a popularity-based approach. For example, the MAP$_L$@100 for *like* tracks is 0.067 for the system that inputs all listening events, whereas it is only 0.014 for a system based on popularity.

Concerning MAP$_E$@k relative to the tracks listened to, the best results are obtained when all the tracks listened by users are the inputs: at rank 100, MAP$_E$@100 is 0.21373 whereas MAP$_S$@100 is only 0.15701 when the inputs are only listening events whose duration is more than 30 seconds (e.g. streams). This result appears logical, because the information is more complete than the other inputs, which are filtered and therefore reduced. Therefore, the identification of neighbours implies more recommendations. More recommendations corresponding to the actual listening events are thus possible. On the other hand, as observed previously, $MAP_L@k$ for tracks *like* is quite low compared to $MAP_E@k$ for tracks played : 0.08581 instead of 0.48216, for example at rank 10. This significant difference corroborates the observation previously done that only 22% of the recommended tracks are implicit *likes*. Regarding recommendations that may be considered as wrong picks, MAP@k for skips or tracks *dislike* are quite high. Indeed, nearly 40% of recommended tracks are skipped, and 21% are not appreciated by users.

| Algorithms | MAP@10 | MAP@100 | MAP@500 |
|---|---|---|---|
| $MAP_E@k$ (events) | | | |
| Popularity | 0.15307 | 0.05102 | 0.03891 |
| UB / events | **0.48216** | **0.21373** | **0.16305** |
| UB / streams | 0.42350 | 0.18064 | 0.14026 |
| UB / likes | 0.20715 | 0.08025 | 0.05656 |
| $MAP_S@k$ (streams) | | | |
| Popularity | 0.10815 | 0.03494 | 0.03294 |
| UB / events | 0.35763 | 0.15701 | **0.13937** |
| UB / streams | **0.37102** | **0.15768** | 0.13927 |
| UB / likes | 0.18982 | 0.07419 | 0.05958 |
| $MAP_L@k$ (likes) | | | |
| Popularity | 0.01980 | 0.01391 | 0.01591 |
| UB / events | 0.08581 | 0.06683 | 0.07145 |
| UB / streams | 0.09303 | **0.07142** | **0.07607** |
| UB / likes | **0.09881** | 0.06702 | 0.06957 |
| $MAP_K@k$ (skips) | | | |
| Popularity | 0.06664 | 0.02683 | 0.02746 |
| UB / events | 0.15613 | 0.08504 | 0.08481 |
| UB / streams | 0.11268 | 0.05975 | 0.06245 |
| UB / likes | **0.03549** | **0.02019** | **0.02147** |
| $MAP_D@k$ (dislikes) | | | |
| Popularity | 0.02389 | 0.01205 | 0.01406 |
| UB / events | 0.06864 | 0.04617 | 0.05011 |
| UB / streams | 0.02385 | 0.01972 | 0.02420 |
| UB / likes | **0.00936** | **0.00600** | **0.00694** |

**Table 6.** MAP@k (rangs k=10,100,500) on different criteria (listening events, streams, skips, implicit likes and dislikes, for the same recommendation engine based on neighbourhoods between users, for different inputs (listening events, streams and likes).

By modifying and filtering the input data of the recommender system, an increase of the MAP@k can be observed. For example, with only the tracks *like* as input, the $MAP_L@10$ relative to the tracks *like* is 0.099 instead of 0.086 for the $MAP_E@10$. The number of appreciated tracks within the recommended tracks is increased here by 13% and these tracks are best ranked in the list of recommendations. This difference appears numerically limited but may be of great importance for some applications.

On the other hand, the differences are more important for skips or tracks *dislike*. Still considering the tracks *like* as inputs, $MAP_D@10$ relative to the *dislike* tracks is lower, 0.009 instead of 0.069 for the $MAP_E@$ 10. The number of unappreciated tracks here is almost divided by 5. Filtering input data significantly reduces the possibilities that the user is recommended a track that he will not appreciate.

Several conclusions can be drawn from this comparisons: by filtering the input data of the recommender systems, the recommendation of tracks *like* is slightly improved, and the recommendation of skipped tracks or tracks *dislike* is quite sharply reduced. Some users may be very sensitive to the recommendation of tracks they do not like. Filtering data over duration and replays may be a good approach to limit the possibility of recommending such tracks.

### 3.4 Filtering lists of tracks

The positive results of the previous section enable consider other applications related to the recommendation, such as the possibility of filtering or reranking the tracks of a list (playlist, radio), taking into account the previous interactions of the user, in particular durations and replays. This problem is related to question 2 in Section 1.

### 3.4.1 General method

Figure 3 proposes an illustration of the approach proposed. From the listening events (durations and replays) of a user, implicit ratings can be estimated for the list of tracks considered. These ratings can then be used to select, filter or rerank the initial list of tracks to better match the tastes of the user.

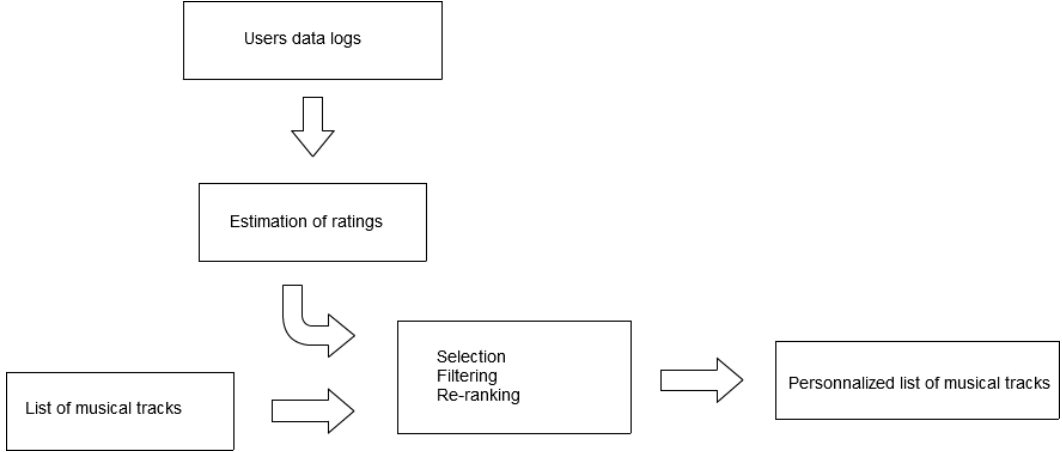

**Figure 3.** Illustration of the filtering process proposed: a list of tracks is filtered regarding the user usage data in order to obtain a list of personalized tracks.

Research in the music recommendation area provides many approaches to estimate the ratings of tracks that a user does not know[3]. Evaluations of these estimations have been the objective of the Netflix prize[23] and are mainly based on the measurement of the error between the real value and the estimated one, for example by considering Root Mean Square Error (RMSE) calculation. The most accurate methods are based on latent factor models[33]. These methods prove to be less efficient on retrieval metrics such as MAP@k, as shown by the results presented in table 5. But the problem raised here is to treat a posteriori a list of tracks, possibly computed by another recommendation engine or by editors. Methods based on latent factor appear thus to be a good choice. Among these methods, the SVD method available in the Apache Mahout toolkit is chosen for the following experiments, with matrix factorization computed by applying Stochastic Gradient Descent algorithm. Settings are fixed empirically: 500 latent factors, 20 iterations and 0.05 as regularization parameter.

### 3.4.2 Score function

The difficulty in applying such model-based systems comes from the fact that there is no explicit rating available of music tracks, on the contrary of the Netflix prize for example. It is therefore necessary to propose a mapping function, denoted by $f$, which transforms the listening data, including durations and replays, into integers or real numbers corresponding to ratings:

$$f : E(u,i) \longrightarrow r(u,i) \qquad (11)$$

where $E(u,i) = \{E(u,i,t), \forall t\}$ is the set of listening events of the track $i$ by the user $u$ and $r(u,i)$ is the implicit rating of the track $i$ by the user $u$, corresponding to an integer ($r(u,i) \in \mathbb{N}$).

Obviously, there is a multitude of possibilities for defining this function $f$, and thus a multitude of possible choices to define what *a user loves a track* means for duration or replay data. The analysis of listening data presented in Section 2.2 indicates that the behaviour of each user can differ greatly from this point of view. Indeed, some users will skip several times the same track, while continuing to love it and expecting to listen to it later. On the contrary, other users, after skipping a track, will never listen to it again. Behaviours are multiple and their study and modelling would require specific work beyond the scope of this study.

In the following, it was chosen empirically 3 simple functions to calculate ratings from listening durations $d(u,i,t)$ and play counts $P(u,i)$. Implicit ratings computed are reduced to known explicit rating schemes and correspond to integers between 1 and 5, using the principle of star rating systems such as Netflix. The objective here is not to find the optimal function, but to verify the contribution of the filtering approach.

Function $f_1$ considers only the replays $P(u,i)$ by setting a threshold at 2 streams. If there is more than 2 streams, it is considered that the track is appreciated and a score between 3 and 5 is assigned, depending on the number of plays. If there is exactly 2 streams, the implicit rating is assumed as more ambiguous and the score is only 3. Finally, if there is only 1 stream or less (only listening events whose duration are less than 30 seconds for example), the score assigned takes a low value or the minimum value of 1 if there is no stream at all. Thus, function $f_1$ is described in equation 12 below:

$$f_1 : E(u,i) \longrightarrow \begin{cases} r(u,i) = 5 & \text{if } P(u,i) \geq 4, \\ r(u,i) = 4 & \text{if } P(u,i) = 3, \\ r(u,i) = 3 & \text{if } P(u,i) = 2, \\ r(u,i) = 2 & \text{if } P(u,i) = 1, \\ r(u,i) = 1 & \text{if } P(u,i) = 0. \end{cases} \quad (12)$$

Functions $f_2$ and $f_3$ consider both durations and replays. Function $f_2$ is based on the implicit *like* and *dislike* definitions introduced in Section 2.2. Thus, if user $u$ implicitly shows that he likes track $i$, that is $L(u,i) = 1$, then the rating assigned is the maximum value of 5. If, on the contrary, user implicitly shows that he/she does not appreciate it, that is, $D(u,i) = 1$, then the rating assigned is the minimum value of 1. Apart from these two cases, the rating assigned is a neutral score of 3. This function $f_2$ is described in Equation 13 below:

$$f_2 : E(u,i) \longrightarrow \begin{cases} r(u,i) = 5 & \text{if } L(u,i) = 1, \\ r(u,i) = 3 & \text{if } D(u,i) = 0 \text{ and } L(u,i) = 0, \\ r(u,i) = 1 & \text{if } D(u,i) = 1. \end{cases} \quad (13)$$

Finally, function $f_3$ considers the number of skips $K(u,i)$ with respect to the number of plays $P(u,i)$. If the number of plays is greater, the rating is globally positive by being greater than or equal to 3, depending on the number of plays. On the other hand, if the number of skips exceeds the number of plays, the rating is less positive, and then changes according to the number of plays. This function $f_3$ is described in Equation 14 below:

$$f_3 : E(u,i) \longrightarrow \begin{cases} r(u,i) = 5 & \text{if } P(u,i) \geq 4 \text{ and } K(u,i) < P(u,i), \\ r(u,i) = 4 & \text{if } P(u,i) \geq 2 \text{ and } K(u,i) < P(u,i), \\ r(u,i) = 3 & \text{if } K(u,i) < P(u,i), \\ r(u,i) = 2 & \text{if } K(u,i) \geq P(u,i) \text{ and } P(u,i) > 0, \\ r(u,i) = 1 & \text{if } P(u,i) = 0. \end{cases} \quad (14)$$

From the user listening data, the functions proposed are applied to compute the implicit ratings for each track listened to. Thus, a list of triplets $(u,i,r(u,i))$ is obtained, with $u \in U$ and $i \in I$. Table 7 shows the distributions of the implicit ratings obtained, according to the different mapping functions, for the data $A$ and $B_V$ (non-hidden part).

|  | Data $A$ | | | | |
| --- | --- | --- | --- | --- | --- |
|  | rating 1 | rating 2 | rating 3 | rating 4 | rating 5 |
| $f_1$ | 17,062,682 / 31.4% | 26,763,189 / 49.2% | 5,555,319 / 10.2% | 2,041,160 / 3.8% | 2,968,370 / 5.5% |
| $f_2$ | 17,062,682 / 31.4% |  | 30,110,215 / 55.4% |  | 7,217,823 / 13.3% |
| $f_3$ | 17,062,682 / 31.4% | 4,224,161 / 7.8% | 23,288,413 / 42.8% | 6,967,469 / 12.8% | 2,847,995 / 5.2% |
|  | Data $B_V$ (non-hidden part of data $B$) | | | | |
|  | rating 1 | rating 2 | rating 3 | rating 4 | rating 5 |
| $f_1$ | 4,584,936 / 29.7% | 6,753,897 / 43.7% | 1,779,249 / 11.5% | 776,519 / 5.0% | 1,559,213 / 10.1% |
| $f_2$ | 4,584,936 / 29.7% |  | 8,495,411 / 55.0% |  | 2,373,467 / 15.4% |
| $f_3$ | 4,584,936 / 29.7% | 1,616,872 / 10.5% | 5,588,008 / 36.2% | 2,218,210 / 14.4% | 1,445,788 / 9.3% |

**Table 7.** Distribution of ratings for data $A$ and $B_V$.

In this table, it is important to note that the number of ratings $r(u,i)$ with the minimum value of 1 is the same regardless of the chosen score function (31.4% of scores for data $A$). This observation is justified by the choices of the three functions tested, each of which considers a track on which the user has performed a short play (less than 30 seconds) and has never listened to. This criterion also corresponds to the implicit *dislike* definition. For the other scores, the distributions are different with a higher density on the rating 2 for the function $f_1$ (49.2% for the data $A$), and on the rating 3 for the function $f_3$ (55.4% for data $A$). More generally, the distributions for the data $A$ are very close to those of the data $B_V$.

The obtained triplets $(u,i,r(u,i))$ constitute the input of the recommender system allowing to estimate the ratings of all the couples $(u,i)$, in particular the couples for which there was no interaction of the user $u$ on the track $i$. The estimated ratings are denoted by $\tilde{r}(u,i)$.

### 3.4.3 Classification of recommended tracks

Estimated ratings should discriminate tracks that the user may like from those that the user may not like. This discrimination corresponds to a binary classification based on estimated ratings. To perform this classification, a Bayesian naive method is applied. The evaluation proposed here is based on the filtering of tracks recommended by the UB recommender system based on user neighbourhoods. This system produces a list of recommended tracks for each user. This list thus constitutes the input of the filtering system experimented. The goal of the filtering process is to increase the proportion of recommended tracks the user will appreciate (*like*) and reduce the proportion of recommended tracks the user will not appreciate (*dislike*). The classification induced is expected to label the tracks by one of the classes *like* or *dislike*.

Figure 4 shows the distributions of the ratings obtained by the function $f_3$ (for example) and estimated for tracks from data *B*, labelled by the classes *like* or *dislike*, as well as the associated Gaussian modelling. For example, in this configuration, distributions allow for possible discrimination with a precision of 80% for a 83% recall. The applied naive Bayes classification allows us to assign a label *like* or *dislike* to each recommended track. For a user $u$ and a track $i$, this is equivalent to obtain estimations of the two functions $D(u,i)$ and $L(u,i)$ which are denoted by $\widetilde{D}(u,i)$ and $\widetilde{L}(u,i)$ in the following.

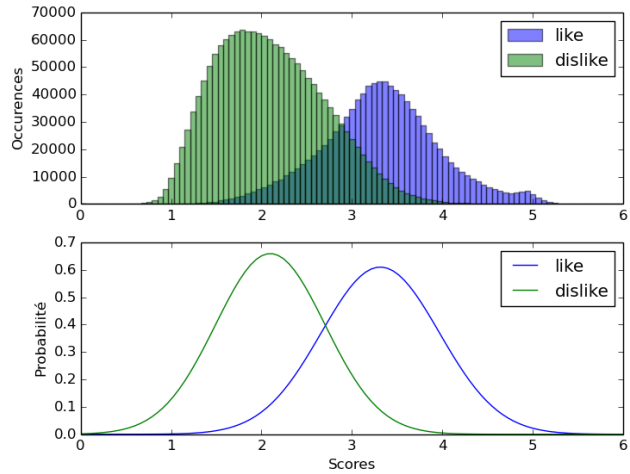

**Figure 4.** Distribution of the two classes like/dislike (implicit) for the ratings estimated by the SVD recommendation engine on the data *B*, according to the mapping function $f_3$ defined by Equation 14.

### 3.4.4 Filtering algorithms

The next step consists in processing the list of recommended tracks in order to improve it, by considering estimated ratings and/or labels. Before any filtering process, this list of tracks recommended for user $u$ is denoted by $R(u)$.

Several algorithms are possible, but only three possibilities are compared here:

- The first algorithm consists in reranking the tracks $i$ of the recommendation list $R(u)$ with respect to the estimated evaluation ratings $\tilde{r}(u,i)$, in descending order. This algorithm, denoted RANK, is described below.

**Algorithm 1** Algorithm RANK

Ranking tracks $i$ of the list $R(u)$ according to $\tilde{r}(u,i)(in descending order)$

- The second algorithm proposed is based on the deletion of tracks that are assumed to be unappreciated by the user. A track assumed to be an implicit *dislike* must therefore be removed from the list of recommended tracks. The algorithm DEL described below consists therefore in browsing the list of tracks recommended for the user $u$, and in removing from this list the tracks $i$ for which $\widetilde{D}(u,i) = 1$.

- The third algorithm proposed relies on the principle of the previous DEL algorithm by removing from the list, the recommended estimated as implicit *dislike*, e.g. tracks for which $\widetilde{D}(u,i) = 1$. The difference with DEL algorithm is the replacement of the track deleted by the first track of the rest of the list for which the estimated score $\tilde{r}(u,i)$ is greater than a empirically fixed threshold $\alpha$. The proposed algorithm named SWAP is described below in the algorithm 3.

**Algorithm 2** Algorithm DEL
>  **for** track $i$ in list $R(u)$ **do**
>    **if** $\widetilde{D}(u,i) = 1$ **then**
>      delete $i$ from list $R(u)$
>    **end if**
>  **end for**

**Algorithm 3** Algorithm SWAP
>  **Input:** R(u), threshold $\alpha$, functions $\tilde{r}$ et $\widetilde{D}$
>  **for** track $i$ in list $R(u)$ **do**
>    **if** $\widetilde{D}(u,i) = 1$ **then**
>      delete $i$ from list $R(u)$
>      **for** track $j$ in list $R(u)$ from the rank of the track $i$ **do**
>        **if** $\tilde{r}(u,j) \geq \alpha$ **then**
>          insert $j$ instead of $i$ in list $\widetilde{R}(u)$
>        **end if**
>      **end for**
>    **else**
>      insert $i$ in list $\widetilde{R}(u)$
>    **end if**
>  **end for**
>  **return** $\widetilde{R}(u)$

### 3.4.5 Evaluation of filtering process

Table 8 shows the distribution of the recommended tracks according to the different criteria used in Section 2.2, for different filtering algorithms and different mapping functions. These distributions are compared to the list of recommended tracks without any filtering process.

| Filtering | Rating | Events | % Streams (> 30s) | % Like | % Skips | % Dislike |
|---|---|---|---|---|---|---|
| No filtering | | **492,898** | 79.0 | 22.1 | 39.1 | 21.0 |
| SWAP | 1 | 373,233 | 87.6 | 28.1 | 33.2 | 12.4 |
| DEL | 1 | 375,911 | 91.7 | 29.4 | **30.8** | 8.3 |
| RANK | 1 | 224,885 | **93.0** | **36.1** | 36.7 | **7.0** |

**Table 8.** Influence of different filtering algorithms on the quality of recommendation at rank 10.

The results obtained correspond to the fixed objective, since the application of the filtering algorithms generally allow to increase the proportion of streams and *like*, while significantly decreasing the proportion of skips and *dislike*. For example, the algorithm DEL applied with mapping function $f_3$ results in more than 29% of tracks *like* instead of 22% without any filtering process. At the same time, the proportion of tracks *dislike* is 8.3% instead of 21% without filter. The comparison between the filtering algorithms indicates that the SWAP algorithm is less accurate than the DEL or RANK algorithms. The explanation for this lower contribution probably comes from the cumulative classification errors, replacing tracks estimated to be *dislike* by tracks estimated to be *like*. This aggregation of errors is not carried out either with the RANK algorithm or with the DEL algorithm, which only take into account the *dislike* tracks. The RANK algorithm has the disadvantage of filtering many tracks, which has an impact on the total number of *like* tracks recommended, although the percentage is much improved (36.1% instead of 22.1% without any filtering process).

It is important to note that the number of recommended tracks actually listened to is falling sharply as expected, since this number includes all tracks, whether liked or not, in the first 10 recommendations. So, the number of tracks *like* is quite close: for example approximately 109,000 without any filtering process, while 110,500 tracks will be *like*d after filtering in the top 10. The difference is important on tracks which are not appreciated : 31,200 tracks *dislike* are present in the top 10 after filtering, while they are over 103,500 before. In this case, filtering process results in the removal of two-thirds of the tracks *dislike* in the top 10. Despite the inevitable errors on classification of tracks *like* and *dislike*, filtering process overall improves the selection of tracks from a pre-established list. The choice between the different settings of the different algorithms can be made according to the desired objective, which can be, for example, to minimize the tracks *dislike* for a specific user, or to reduce the tracks *dislike* while preserving the maximum number of tracks *like* for a less demanding user who usually listens to a lot of recommended tracks.

### *3.4.6 Evaluation of the recommender system after filtering*

Another way of evaluating these filtering operations is to compare the MAP at different ranks, on the same principle proposed in Section 3.3. It is expected that the filtering process will have a positive influence on the results given in Table 6. Table 9 shows MAP at ranks 10, 100, and 500 obtained by filtering the tracks recommended by user-based recommender system, for different types of inputs according to the LED, RANK, and SWAP algorithms.

The results show that MAP for listening events and streams remain higher without any filtering process. This observation was expected since the filtering process consists in deleting tracks. The number of tracks in the list that are then listened to decreases accordingly. Even if deleted tracks are unappreciated tracks in the list, such as tracks *dislike*, the $MAP_E@k$ and $MAP_S@k$ evaluations essentially quantify the importance of the filtering process carried out. At first glance, this decrease appears to be a negative result, which is not the case, since the objective here is mainly to evaluate the variations on the tracks *like* or *dislike*.

The filtering process applied has a very low impact on $MAP_L@k$ relative to tracks *like*: it sometimes results in slightly better $MAP_L@10$ (0.0867 for DEL instead of 0.08581 without any filtering process), but sometimes worse at higher ranks (0.06451 for DEL instead of 0.06683 without any filtering process, at rank 100). The algorithm SWAP is no better than the algorithm DEL, which is probably due to a lack of precision on the classification of tracks into *like* / *dislike* groups. The algorithm RANK allows to obtain a lower $MAP_L@k$ score than for the algorithms DEL or SWAP : tracks with the highest estimated ratings are not always the tracks that have been listened to and appreciated by users.

On the other hand, the difference is greater on the $MAP_K@k$ and $MAP_L@k$, relative to skips and tracks *like*. For example, the $MAP_K@k$ relative to the skips drops significantly for all the algorithms tested (for example 0.03732 for DEL instead of 0.08504 at rank 100 without any filtering process). Similarly, the $MAP_D@k$ relative to the tracks *dislike* drops very sharply as well: 0.00932 for DEL instead of 0.04617 at rank 100 without any filtering process. This decrease in MAP confirms the interest of filtering to remove from the list the tracks that the user would not appreciate. The algorithm RANK obtains the lowest $MAP_K@k$ and $MAP_L@k$, in particular with the function $f_2$. This algorithm, by placing the tracks with the lowest ratings estimated at the end of the list, proves to be quite effective in this context. The balance between $MAP_L@k$ and $MAP_D@k$, relative to the tracks *like* and *dislike*, has of course to be adjusted according to the applications concerned, as discussed above.

In general, the results are better for the rating function $f_3$, which takes into account both replays and the listening event durations. Thus, it is certainly necessary to consider other parameters of description of the usage of the users, which could further improve the efficiency of the filtering process.

These experiments confirm the interest of the filtering algorithms proposed, based on durations and replays. The list of tracks passed as input (e.g. the list of recommended tracks here) can be filtered and thus greatly reduce skips and tracks *dislike* while preserving tracks *like*.

## 4 Conclusions and Perspectives

The questions raised in this article concern the consideration of replay and listening duration information. The dataset collected on a streaming platform allows to show that replays and listening durations can be effectively considered to improve the qualities of music recommender systems. In a first step, the comparison of the recommendation engines confirmed the results of previous works: the best accuracy is obtained by neighbourhood-based approaches and by taking into account binary usage information. Play counts do not have a positive influence in this case.

There are two main contributions in the studies proposed. The first one concern the possibility of filtering the input data of recommender systems with respect to the listening event durations and replays, in order to improve the quality of the recommendations. In particular, it is thus possible to greatly attenuate the number of recommended tracks which will not be appreciated by users. The second contribution is the advantage of estimating implicit user ratings based on replays and listening durations, in order to estimate implicit ratings for all tracks and thus post-filter lists of tracks to adapt them to each user. The experiments show in particular that the number of tracks that the user will not appreciate can thus be greatly diminished. There are many applications for the improvement of radios or playlists by adapting them to each user and thus preventing users from changing radio or playlist they listen to.

However, the proposed evaluations have limitations, particularly as they are offline. They are therefore limited to the a posteriori analysis of user usages. Implementation of online evaluations would allow to verify the results presented here, in particular the advantage of filtering recommendation lists based on implicit ratings.

There are many perspectives. It would be important to consider the relevance of other types of implicit ratings based on user actions such as collecting (favourites, playlists) for example. Other more explicit elements should also be tested such as the *thumb-up/thumb-down* often offered while listening to a track.

As mentioned earlier, the limitations of the approach proposed concern the estimation of implicit ratings based on listening events. Each user has its own behaviours, so the main challenge is to be able to propose implicit rating processes specific to each user and to evaluate the overall influence on the quality of the recommendation.

## Acknowledgements

The author wishes to thank Simbals team, Deezer R&D and Recommendation teams for making this work possible, in particular Manuel Moussalam, Thomas Bouabca and Aurélien Hérault.

| Algorithms | Rating functions | MAP@10 | MAP@100 | MAP@500 |
|---|---|---|---|---|
| $MAP_E$@k (events) | | | | |
| No | | **0.48216** | **0.21373** | **0.16305** |
| SWAP | $f_1$ | 0.28240 | 0.12461 | 0.06762 |
| SWAP | $f_2$ | 0.36233 | 0.16090 | 0.09064 |
| SWAP | $f_3$ | 0.34120 | 0.14910 | 0.08390 |
| DEL | $f_1$ | 0.31897 | 0.10596 | 0.06885 |
| DEL | $f_2$ | 0.31678 | 0.12081 | 0.08945 |
| DEL | $f_3$ | 0.36334 | 0.13010 | 0.09316 |
| RANK | $f_1$ | 0.18153 | 0.06158 | 0.05048 |
| RANK | $f_2$ | 0.05949 | 0.03502 | 0.05268 |
| RANK | $f_3$ | 0.18230 | 0.07031 | 0.07131 |
| $MAP_S$@k (streams) | | | | |
| No | | **0.35763** | **0.15701** | **0.13937** |
| SWAP | $f_1$ | 0.23471 | 0.09941 | 0.06663 |
| SWAP | $f_2$ | 0.28517 | 0.012207 | 0.08343 |
| SWAP | $f_3$ | 0.29221 | 0.12360 | 0.08411 |
| DEL | $f_1$ | 0.27621 | 0.09642 | 0.07323 |
| DEL | $f_2$ | 0.27813 | 0.10938 | 0.08892 |
| DEL | $f_3$ | 0.32437 | 0.12294 | 0.09877 |
| RANK | $f_1$ | 0.16864 | 0.05766 | 0.05466 |
| RANK | $f_2$ | 0.04928 | 0.02779 | 0.04521 |
| RANK | $f_3$ | 0.16864 | 0.06527 | 0.07023 |
| $MAP_L$@k (likes) | | | | |
| No | | 0.08581 | **0.06683** | **0.07145** |
| SWAP | $f_1$ | 0.07161 | 0.05370 | 0.05024 |
| SWAP | $f_2$ | 0.07908 | 0.05960 | 0.05615 |
| SWAP | $f_3$ | 0.08150 | 0.06213 | 0.05867 |
| DEL | $f_1$ | 0.08193 | 0.05679 | 0.05781 |
| DEL | $f_2$ | 0.07958 | 0.05731 | 0.06010 |
| DEL | $f_3$ | **0.08670** | 0.06451 | 0.06721 |
| RANK | $f_1$ | 0.05404 | 0.03744 | 0.04180 |
| RANK | $f_2$ | 0.01900 | 0.01630 | 0.02437 |
| RANK | $f_3$ | 0.05938 | 0.04347 | 0.04959 |
| $MAP_K$@k (skips) | | | | |
| No | | 0.15613 | 0.08504 | 0.08481 |
| SWAP | $f_1$ | 0.09522 | 0.05031 | 0.04109 |
| SWAP | $f_2$ | 0.10463 | 0.05880 | 0.04723 |
| SWAP | $f_3$ | 0.08831 | 0.04910 | 0.03966 |
| DEL | $f_1$ | 0.10233 | 0.04159 | 0.03936 |
| DEL | $f_2$ | 0.06739 | 0.03393 | 0.03549 |
| DEL | $f_3$ | 0.08817 | 0.03732 | 0.03709 |
| RANK | $f_1$ | 0.06851 | 0.02882 | 0.03169 |
| RANK | $f_2$ | **0.01577** | **0.01180** | **0.02376** |
| RANK | $f_3$ | 0.05473 | 0.02534 | 0.03498 |
| $MAP_D$@k (dislikes) | | | | |
| No | | 0.06864 | 0.04617 | 0.05011 |
| SWAP | $f_1$ | 0.02377 | 0.01788 | 0.01438 |
| SWAP | $f_2$ | 0.04209 | 0.02862 | 0.02350 |
| SWAP | $f_3$ | 0.02493 | 0.01885 | 0.01488 |
| DEL | $f_1$ | 0.01717 | 0.00845 | 0.00877 |
| DEL | $f_2$ | 0.01727 | 0.01211 | 0.01352 |
| DEL | $f_3$ | 0.01556 | 0.00932 | 0.01013 |
| RANK | $f_1$ | 0.00635 | 0.00461 | 0.01279 |
| RANK | $f_2$ | **0.00554** | **0.00343** | **0.00569** |
| RANK | $f_3$ | 0.00614 | 0.00402 | 0.011749 |

**Table 9.** MAP@k (k=10,100,500) on different criteria (listening events, streams, skips, implicit like and dislike, for the different filtering algorithms DEL, SWAP and RANK.